\documentclass[useAMS,usenatbib]{mn2e}
\usepackage{times}
\usepackage{graphicx, latexsym, amssymb, amscd, psfrag}
\usepackage{epsfig}

\title[The Mass Assembly of Fossil Groups of Galaxies in the Millennium Simulation]{The Mass Assembly of Fossil Groups of Galaxies in the Millennium Simulation}
\author[Dariush et al.]{Ali Dariush\thanks{E-mail:
aad@star.sr.bham.ac.uk}$^{1}$,
Habib G. Khosroshahi$^{1}$, Trevor J. Ponman$^{1}$, Frazer Pearce$^{2}$, 
\newauthor Somak Raychaudhury$^{1}$ \&  
Will Hartley$^{2}$\\
$^{1}$School of Physics and Astronomy, 
University of Birmingham, Birmingham B15~2TT, UK\\
$^{2}$School of Physics and Astronomy,  
University of Nottingham, Nottingham, NG7~2RD, UK}

\begin{document}

\date{}

\pagerange{\pageref{firstpage}--\pageref{lastpage}} \pubyear{2006}

\maketitle

\label{firstpage}
\begin{abstract}

The evolution of present-day fossil galaxy groups is studied in the
Millennium Simulation. Using the corresponding Millennium gas
simulation and semi-analytic galaxy catalogues, we select fossil
groups at redshift zero according to the conventional observational
criteria, and trace the haloes corresponding to these groups backwards
in time, extracting the associated dark matter, gas and galaxy
properties. The space density of the fossils from this study is
remarkably close to the observed estimates and various possibilities
for the remaining discrepancy are discussed. The fraction of X-ray bright
systems which are fossils appears to be in reasonable agreement with
observation, and the simulations predict that fossil systems
will be found in significant numbers (3-4\% of the population) 
even in quite rich clusters. We find that fossils
assemble a higher fraction of their mass at high redshift, compared
to non-fossil groups, with the ratio
of the currently assembled halo mass to final mass, at any epoch, being
about 10 to 20\% higher for fossils. This
supports the paradigm whereby fossils represent undisturbed,
early-forming systems in which large galaxies have merged
to form a single dominant elliptical. 

\end{abstract}


\section{Introduction}

Galaxy groups are believed to play a key role in the formation and
evolution of structure in the universe as, within a hierarchical
framework, they span the regime between individual galaxies and massive
clusters. They are also more varied in their properties than galaxy 
clusters, as seen when various scaling
relations are compared with those of galaxy clusters
\citep{b91,b136,b171,b3,b121,b4,b178,b89,b177,b90}.  For instance, the
relation between the luminosity and temperature of the X-ray emitting
hot intergalactic medium (the $L\!-\!T$ relation) has a larger
scatter and a different slope for groups, when compared to similar
properties of clusters.  Various feedback mechanisms are often invoked
to explain these differences.  In addition, due to their lower
velocity dispersion, groups are rapidly evolving systems, and galaxy
mergers within groups can have a more significant effect on these
relations than in clusters. In principle, the presence of cool cores
and active galactic nuclei (AGN), as well as the star formation
history, are all affected by major interactions in the heart of a
group or cluster. It would therefore be useful to find a class of
groups or clusters with no major mergers in their recent history, to
provide a baseline for the evolution of a passive system, with no
major disruption.

Fossil groups are good candidates for such a class of objects.  They are
distinguished by a large gap 
between the brightest galaxy and the fainter members, with an
under-abundance of $L_\ast$ galaxies.  \citet{b300} suggest that, for
an X-ray detected group, the merging timescale for the most luminous
group members ($L\approx L_\ast$) is of order of a few tenths of an
Hubble time, in agreement with the numerical simulations. A single
giant elliptical galaxy can form as a result of multiple mergers
within a few Gyr \citep{b15}.  Thus, it is likely that one can
find merged groups in the form of an isolated giant elliptical galaxy
with an extended halo of hot gas, since the
timescale for gas infall is longer than that on which galaxies merge
\citep{b135}.  In such systems, the brighter galaxies, which have a
relatively shorter merging timescale, are expected to merge earlier
leaving the fainter end of the luminosity function intact \citep{b55,b105}.

Following the discovery of a fossil group having the above
characteristics from ROSAT observations \citep{b130}, more fossil
systems have been identified \citep{b120,b170,b70,b140,b95,b65}.  They
are generally based on the definition of fossil groups from
\citet{b65}, i.e. groups with a minimum X-ray luminosity of $L_{\rm
X,bol} \approx 0.25 \times 10^{42} h^{-2}$erg s$^{-1}$, as well as
minimum magnitude difference of two between  the first and
second ranked galaxies, within half the projected radius that encloses
an overdensity of 200 times the mean density of the universe ($R_{200}$).  
For an NFW
profile \citep{b123}, this is roughly equivalent to $R_{500}$, the
radius enclosing an overdensity of 500 times the mean (for NFW haloes
of the appropriate concentration, $R_{500} \sim 0.59 \times
R_{200}$). A few of these fossil groups have been the subject of detailed
investigations \citep{b75,b185,b165,b167,b45,b100,b80}. 

While most
previous studies have focused on X-ray properties of fossils, there is
also emerging evidence that the galaxy properties in fossils are
different from those in non-fossils \citep{b82}. For instance the 
isophotal shapes of the central fossil galaxies appear to be non-boxy, 
suggesting that they may have formed in gas rich mergers. 
Various observational and
theoretical studies have suggested a significant fraction of galaxy
groups to be fossils \citep{b170,b65,b50,b115,b147}, though often the
criteria used to define fossils in theoretical work are not easy to
relate to observational studies.

Fossils may represent extreme examples of a continuum of group properties
-- they are consistently found to be outliers in the usual scaling
relations involving optical, X-ray and dynamical properties \citep{b85}. 
While fossils fall on the L-T relation of non-fossil groups and clusters, 
they appear to be both hotter and more X-ray luminous than non-fossils of 
the same mass. Cooler fossil groups
also show lower entropy than their non-fossil counterparts. According to 
\citet{b85}, the haloes of fossil groups appear to be more concentrated than 
those of non-fossil systems, for a given mass,
which suggests that fossils have an early formation epoch. 
As such, we have much to learn from them, and the investigation of objects
with similar properties in cosmological simulations can provide important
insights into the physical processes that underly the scaling relations. It
can also reveal limitations in the numerical simulations, related to the
treatment of physical effects like pre-heating, feedback and merging, which
are difficult to model.  It is thus important to study the formation and
evolution of these systems in the cosmological N-Body simulations which
have become essential tools for studying formation of large scale structure
in the Universe.

In this paper we use the Millennium simulation \citep{b160} 
 together with the semi-analytic
models \citep{b40} of galaxy formation within dark matter haloes and
the Millennium gas simulation \citep{b128}, to identify fossil groups,
study their properties in the simulations and make a comparison to the
observations. We begin with a brief discussion in \S2 of the Millennium
Simulation, and the implemented semi-analytic galaxy catalogues and
gas simulations.  In \S3 we discuss our method of identifying {\it optical} and {\it X-ray} fossil
groups from these catalogues.  In \S4, we discuss the various
properties of these fossil groups, their abundance in the local
Universe and the evolution of simulated X-ray fossils with time. Finally, in \S5, we summarize
the implications of our results in terms of the evolution of fossil
groups in the context of multiwavelength observations. 
Throughout the paper we adopt $H_{0} = 100 \,h$ km s$^{-1}$ Mpc$^{-1}$ for the Hubble constant.


\section[]{Description of the Simulations}

In order to extract fossil groups in the Millennium simulation, using observational selection
criteria, we require a simulation suite that includes the baryonic
physics of hot gas and galaxies, as well as a high resolution dark
matter framework and a sufficient spatial volume to limit the effects
of cosmic variance. For this study we use the dark matter Millennium
Simulation \citep{b160}, a 10-billion particle model of a comoving
volume of side 500$h^{-1}$ Mpc, on top of which a publicly available
semi-analytic galaxy model \citep{b40} has been constructed. For the
hot gas we have repeated the Millennium simulation with a lower
resolution simulation including gas physics utilising the same volume,
phases and amplitudes as the original dark-matter-only model. This run
accurately reproduces the structural framework of the Millennium
Simulation \citep{b128}. Below we summarize the main characteristics
of the above simulations.

\subsection{The Millennium Simulation}

The Millennium Simulation is based on a Cold Dark Matter cosmological
model of structure formation, with a Dark Energy field $\Lambda$. The
basic assumptions are those of an inflationary universe, dominated by dark
matter particles, leading to a bottom-up hierarchy of structure
formation, via collapsing and merging of small dense haloes at high
redshifts, into the large virialised systems such as groups and
clusters that contain the galaxies that we observe today.  The
simulation was performed using the publicly available parallel
TreePM code Gadget2 \citep{b155}, achieving a 3D dynamic range of
$10^5$ by evolving 2160$^3$ particles of individual mass
$8.6\times10^{8}h^{-1}$ M$_{\odot}$, within a co-moving periodic box of
side 500$h^{-1}$ Mpc, and employing a gravitational softening of
5$h^{-1}$ kpc, from redshift $z=127$ to the present day.  The
cosmological parameters for the Millennium Simulation were:
$\Omega_\Lambda = 0.75, \Omega_M = 0.25, \Omega_b = 0.045, h = 0.73, n
= 1$, and $\sigma_8 = 0.9$, where the Hubble constant is characterised
as $100 \,h \,{\rm km s^{-1} Mpc^{-1}}.$ These cosmological parameters are
consistent with recent combined analysis from {\it WMAP} data \citep{b146}
 and the 2dF galaxy redshift survey \citep{b31},
although the value for $\sigma_8$ is a little higher than would
perhaps have been desirable in retrospect.

The derived dark matter halo catalogues include haloes down to a
resolution limit of 20 particles, which yields a minimum halo mass of
1.72$\times 10^{10}h^{-1}$ M$_{\odot}$. Haloes in the simulation are
found using a friends-of-friends (FOF) group finder, tuned to 
extract haloes with overdensities of at least 200 relative to the
critical density.  Within a FOF
halo, substructures or subhaloes are identified using the SUBFIND
algorithm developed by \citet{b155}, and the treatment of the orbital 
decay of satellites is described in the next section.

During the Millennium Simulation,
64 time-slices of the locations and velocities of all the particles
were stored, spread approximately logarithmically in time between
$z=127$ and $z=0$. From these time-slices, merger trees are built by
combining the tables of all haloes found at any given output time, a
process which enables us to trace the growth of haloes and their
subhaloes through time within the simulation.


\subsection{The Semi Analytic model} 
\label{semianalytic}

Using the dark matter haloes of the \citet{b160} simulation,
\citet{b40} have simulated the growth of galaxies, and their central
supermassive black holes, by self-consistently implementing semi-analytic models of
galaxies on the outputs of the Millennium
Simulation. The semi-analytic catalogue contains 9 million galaxies
at $z=0$ down to a limiting absolute magnitude of $M_R\!-\!5
\log\, h = -16.6$, observed in $B$, $V$, $R$, $I$ and $K$ filters. The
models focus on the growth of black holes and AGNs as feedback
sources.  The inclusion of AGN feedback in the semi-analytic model
(which allows the cooling flow to be suppressed in massive haloes that
undergo quasi-static cooling) and its good agreement with the observed
galaxy luminosity function, colour distribution and the clustering
properties of galaxies, make this catalogue well-matched and
suitable for our study of fossil systems.

In the semi-analytic formulation, galaxies initially form within
small dark matter haloes. As the simulation evolves, such a halo may
fall into a larger halo. The semi-analytic galaxy within this halo
then becomes a satellite galaxy within the main halo and follows the
track of its original dark matter halo (now a subhalo) until the
mass of the subhalo drops below 1.72$\times 10^{10}h^{-1}$ M$_{\odot}$, which
corresponds to a 20-particle limit in the Millennium Simulation. At
this point the galaxy is assumed to spiral into the centre, on some
fraction of the dynamical friction timescale, where it merges with the
central galaxy of the larger halo \citep{b40}.

\subsection{The Millennium Gas Simulation}

The Millennium Gas Simulations are a suite of hydrodynamical models,
utilising the same volume, and values of initial perturbation amplitudes and phases
as the parent dark-matter-only Millennium Simulation. Each of the
three models completed to date contains additional baryonic
physics: The first does not follow the effects of radiative cooling
and so overpredicts the luminosities of group-scale objects
significantly. The second includes a simple preheating scheme that is
tuned to match the observed X-ray properties of clusters at the
present day and the third includes a simple feedback model that 
matches the observed properties of clusters today, as well as having
some chance of following the time evolution. We have used the second
of these models in this work, as we only utilise the hydrodynamical
properties of the groups at $z=0$, where the observational and
simulation results are well matched.

Each of the Millennium Gas Simulations consists of $5 \times 10^8$
particles of each species, resulting in a dark matter mass of $1.422
\times 10^{10}h^{-1}$ M$_\odot$ per particle and a gas mass of $3.12
\times 10^{9}h^{-1}$ M$_\odot$ per particle. The Millennium Simulation
has roughly 20 times better mass resolution than this and so some
perturbation of the dark matter halo locations is to be expected. In
practice the position and mass of dark matter haloes above
$10^{13}h^{-1}$ M$_\odot$ are recovered to within $50\,h^{-1}$kpc
between the two volumes, allowing straightforward halo-halo matching
in the large majority of cases. 

The Millennium gas simulations used exactly the same cosmological
parameters as those stated above. With the inclusion of a gaseous
component, additional care needs to be taken in choosing the
gravitational softening length in order to avoid spurious heating
\citep{b162}. We use a comoving value of $25(1+z)h^{-1}$ kpc,
roughly 4\% of the mean interparticle separation \citep{Borgani} until
$z=3$, above which a maximum comoving value of $100h^{-1}$ kpc
pertains. We have adopted a different output strategy for the
Millennium Gas Simulations, preferring to output uniformly in time
with an interval roughly corresponding to the dynamical time of
objects of interest. This strategy results in 160 rather than 64
outputs and places particular emphasis on the late stages of the
simulation. 


\section{Sample Selection}

\subsection{Definition of fossils}

Fossil groups are selected according to a combination of X-ray and
optical criteria, based on the observational definition given by
\citet{b65}, which is widely followed in the literature.  Their X-ray
luminosity must satisfy $L_{\rm X,bol} \geq 0.25 \times 10^{42}
h^{-2}$erg s$^{-1}$, and the difference between the $R$-band
magnitudes of the first and second ranked galaxies, within half the
projected radius enclosing 200 times the mean density ($R_{200}$), must be
$\Delta m_{12}\!\geq\! 2$ magnitudes.  A limit of 0.5$R_{200}$ is
used because $L_\ast$ galaxies within this radius
should spiral into the centre of the group
due to orbital decay by dynamical friction within a Hubble
time \citep{b65}. The limit on the
bolometric X-ray luminosity $L_{\rm X,bol}$ helps to exclude poor
groups and individual galaxies with a few small satellites. Such groups are 
often not in dynamical
equilibrium, and in addition there might be a gap in their galaxy
luminosity function simply as a result of the small numbers of
galaxies involved. We address this issue in \S~\ref{sch}. 

\begin{figure}
\epsfig{file=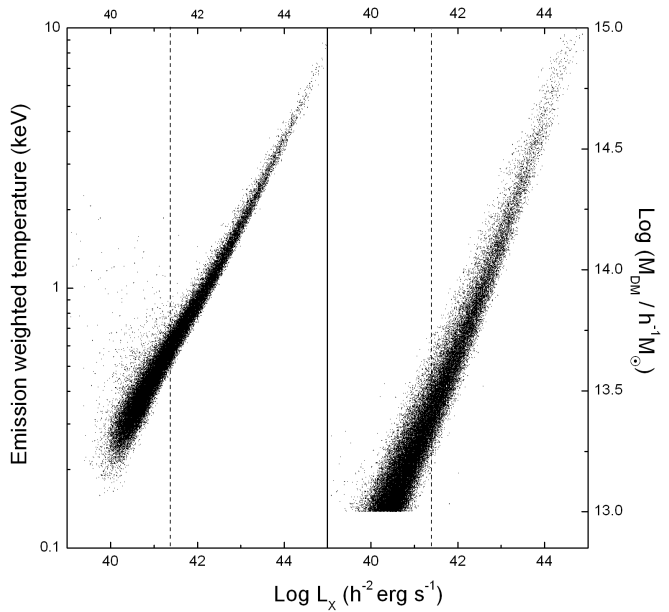,width=3.6in}
\caption{ The bolometric X-ray luminosity versus dark matter halo
temperature ({\it left panel}) and mass within $R_{200}$ ({\it right panel}) for all haloes in the
Millennium gas simulation. The vertical dashed lines correspond to the X-ray luminosity threshold $L_{\rm X,bol} =
0.25\times 10^{42}\,h^{-2}$erg s$^{-1}$ adopted in this paper for defining fossil groups. The cutoff $M(R_{200}) \geq 10^{13}\,
h^{-1}\,$M$_{\odot}$ is adopted in \S~\ref{selection}.}
\label{aadfig1}
\end{figure}


\subsection{Selection of Fossil galaxies}
\label{selection}

For this investigation, we first selected dark matter haloes from the
Millennium Gas Simulation with masses $M(R_{200}) \geq 10^{13}\,
h^{-1}\,$M$_{\odot}$. As Fig.~\ref{aadfig1}  demonstrates, all
haloes for which $L_{X,bol} \!\geq\! 0.25\times 10^{42}\, h^{-2}$erg
s$^{-1}$ are expected to be included in this sample and thus our fossil
sample will be complete. A bolometric X-ray luminosity of $0.25\times
10^{42}\, h^{-2}$ erg s$^{-1}$ corresponds to a temperature of
$T\!\approx\! 0.5$ keV. This constraint on the halo masses results in
an initial sample of 51538 dark matter haloes, within which we search
for fossil groups.

In order to extract  information about dark matter and galaxy properties, 
for each of these 51538 haloes found in the Millennium Gas Simulation, 
the counterpart in the Millennium Simulation needs to be found. This
is a straightforward procedure because the simulated volume  and the
amplitudes and phases of the initial power spectrum were
matched. However, the dark-matter-only Millennium Simulation has 20
times the mass resolution of the corresponding Millennium Gas Simulation
and so contains some additional small scale power. This leads to small
offsets (typically around twice the gravitational softening length) in the
final co-ordinates of equivalent haloes, and even smaller mass
differences (typically less than $5\%$). Of the 51538 haloes, 48774 ($\sim$95\%)
have corresponding haloes
identified in the Millennium Simulation. For each of these matched
haloes we extracted the coordinates and $BVRIK$
magnitudes for each galaxy contained within $R_{200}$, from the
publically available catalogue of \citet{b40}.

\begin{figure}
\epsfig{file=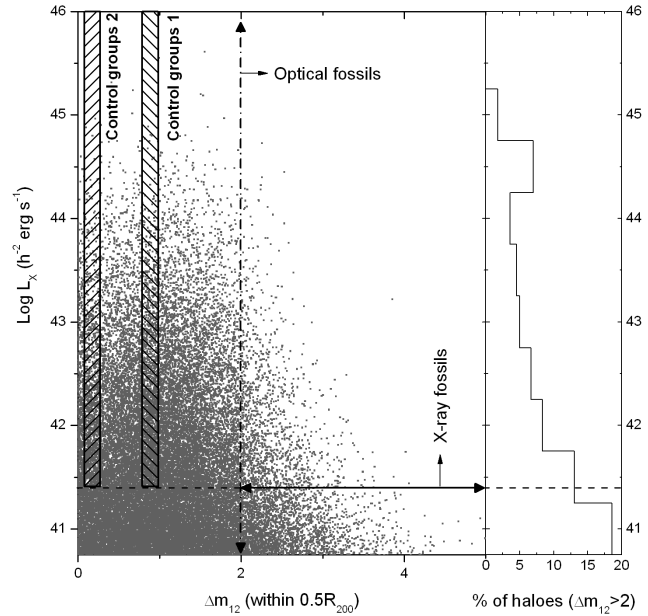,width=3.4in}
\caption{X-ray luminosity versus the $R$-band luminosity gap $\Delta m_{12}$
within 0.5$R_{200}$ for each of
the dark matter haloes with gas properties. The horizontal dashed-line intersects the
vertical axis at  $L_{X,bol} = 0.25\times 10^{42} h^{-2}$erg s$^{-1}$.
 The top-right part of the
graph shows the region for which $\Delta m_{12} \geq 2$ mag and
$L_{X,bol} \geq 0.25\times 10^{42} h^{-2}$erg s$^{-1}$, the optical and X-ray criteria 
that jointly define fossil groups. The two shaded regions show the 
location of our control samples (see \S~\ref{selection} for details). 
The histogram shows the fraction of optical fossils  
in each bin of $L_{X,bol}$.}
\label{aadfig2}
\end{figure}

The simulated properties of the galaxies occupying each dark matter halo
were then used to calculate $\Delta m_{12}$, i.e.  the difference in
$R$-band magnitude of the first and second ranked galaxies within
0.5$R_{200}$ of the centre of the halo. Out of the 48774 matched haloes,
6502 are found to be {\it optical fossil groups}, i.e.  haloes with $\Delta
m_{12}\!\geq\! 2$~mag, among which 1300 are {\it X-ray fossil groups},
i.e. optical fossil groups with $L_{X,bol} \geq 0.25\times 10^{42}\,
h^{-2}$~erg~s$^{-1}$. As can be seen from Fig.~\ref{aadfig2}, X-ray fossil
groups do not form a separate population but are rather extreme examples of
a smooth distribution. It is also clear that the spread in $\Delta m_{12}$
increases dramatically with decreasing in the X-ray luminosity and hence
the enclosed mass, in agreement with the conditional luminosity function
(CLF) formalism of \citet{b187}.

Control groups are necessary for our study in order to allow us to
compare the properties of X-ray fossil and non-fossil groups. We define two
control group samples, based on the magnitude difference
of the two brightest members of the group (within 0.5$R_{200}$ of the 
centre of the dark matter halo): (i) $0.8\! \leq\! \Delta
m_{12}\! \leq\! 1.0$, and (ii) $0.1\! \leq\! \Delta m_{12}\! \leq\!
0.3$.  While members of the former are examples of intermediate groups, the
latter could be regarded as a class of extreme non-fossil groups in
that they contain at least two galaxies of very similar
magnitude. The bolometric X-ray luminosity limit for each of the
control samples is the same as that of the X-ray fossil group sample 
(see Fig.~\ref{aadfig2}).

\subsection{The likelihood of finding groups with $\Delta m_{12} \geq 2$ 
at random}
\label{sch}

One of the central criteria used to define fossils is the absence of
galaxies within a range of two magnitudes of the brightest galaxy
($\Delta m_{12} \geq 2$). However, for groups with only a small number
of members, there is a significant probability of obtaining such a
luminosity gap as a natural consequence of the high-end tail of the
galaxy luminosity distribution.
To quantify the likelihood of obtaining a value of $\Delta m_{12}\!
\geq\! 2$ by chance, \citet{b65} performed 10$^4$ Monte Carlo
simulations for groups and clusters with absolute magnitudes
selected at random from a Schechter function \citep{b145}. Using the
parameters of the composite luminosity function of MKW/AWM clusters
\citep{b180}, they found that for the systems of $\sim$40
galaxies, 0.4$\pm$0.06\% of the generated luminosity functions had
$\Delta m_{12} \!\geq\! 2$.

We performed a similar analysis for groups spanning a range in richness,
and using parameters appropriate to our data from the Millennium simulation.
For twenty classes of groups, containing 
10, 15, 20, 25,... galaxies respectively, we 
randomly generated galaxies
according to a Schechter luminosity function. (None of our X-ray fossil 
groups from the Millennium simulation contain fewer than 10 galaxies.)
The characteristic magnitude M$_\ast \sim -22.1$ and faint-end
exponent $\alpha \sim -1.19$, were adopted from a 
fit of the Schechter function to the $R$-band luminosity function of the 
semi-analytic catalogue. A magnitude cut off of $-17.4$ was then applied 
to the magnitude of generated 
galaxies as this is  the $R$-band magnitude completeness limit of the 
semi-analytic catalogue. 10$^6$ simulations were carried out for each richness class of group.

\begin{figure}
\epsfig{file=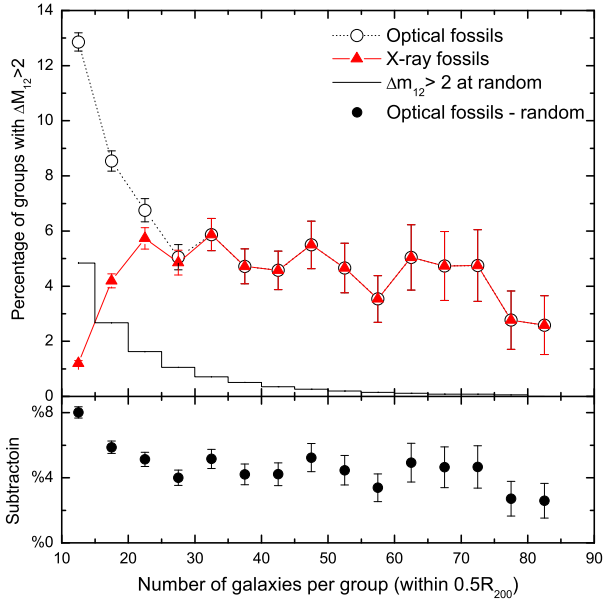,width=3.7in}
\caption{The  histogram shows the incidence rate of $\Delta m_{12} \geq~2$
 occurring by chance from a random population of galaxies selected
from a Schechter luminosity function in comparison to fraction of 
 optical fossils ({\it open circles})
 and X-ray fossils ({\it solid triangles}) in the Millennium simulation,
 as a function of number of galaxies per halo within 0.5$R_{200}$.
 The lower panel plot ({\it filled circles}) shows the result of subtracting
 random groups from the optical fossil groups.}
\label{aadfig3}
\end{figure}

Fig. \ref{aadfig3} compares the percentage of optical and X-ray fossil
groups from the Millennium simulation as a function of number of galaxies
within 0.5$R_{200}$ for each dark matter halo, with those populated using a
Schechter luminosity function as detailed above. The lower panel shows the
result when the expected number of
randomly generated groups with $\Delta m_{12}\! \geq\! 2$, is subtracted from 
the optical fossils. For poor systems, the incidence of `statistical 
fossils' is significant. It can be seen that
approximately one third of the fossil systems with fewer than 25 galaxies
seen in the Millennium data can be attributed to statistical chance (as
opposed to the result of
physical processes generating a non-statistical luminosity
gap). However, even after these random fossils are removed,
the fraction of optical fossils increases as the number 
of galaxies within dark matter haloes decreases. In contrast,
for X-ray fossils, many of the poor haloes which qualify
as 'statistical fossils' fail to pass the X-ray luminosity threshold
criterion, so the chance fraction is never much larger than 20\%.
We return to this issue in Sec.~\ref{abundance}.
For groups with more than 30 galaxies, the fraction of fossils meeting
the $\Delta m_{12}\! \geq\! 2$ criterion by chance drops below 1.0\%,
and soon becomes negligible.

\section{Results}

\subsection{The Luminosity Gap Statistic}

\begin{figure}
\epsfig{file=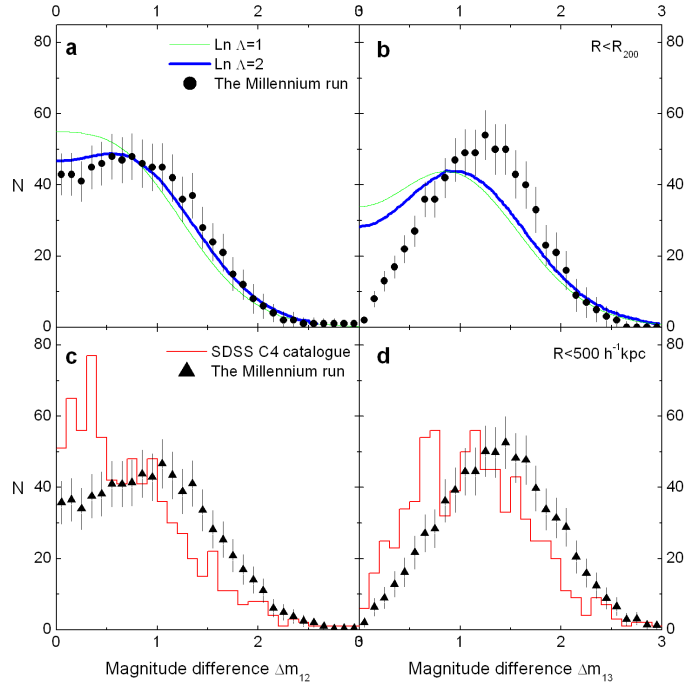,width=3.65in}
\caption[] {The R-band luminosity gap distribution for haloes from the
Millennium semi-analytic model within the
mass range $M=0.5\!-\!10 \times 10^{14}\, h^{-1}\,$M$_{\odot}$,
evaluated relative to the first and second most luminous galaxies
 and the first and the third most luminous galaxies, superposed on the theoretical model of
  \citet{b115} and the SDSS data for the same mass range but different searching radius ($R_{200}$ for the model and 
  projected radius of 500$h^{-1}$kpc for the SDSS data). The Millennium data are plotted
 within $R_{200}$ ({\it closed circles}) as well as the projected radius of 500$h^{-1}$kpc ({\it triangles}).
 (a), (b): The luminosity gap statistic predictions of the theoretical model
of \citet{b115} with $\ln \Lambda=1$ ({\it thin green line}) and $\ln
\Lambda=2$ ({\it thick blue line}). (c), (d): The $r$-band luminosity gap
distribution from 730 clusters ({\it red histogram}) in the SDSS C4
Catalogue of \citet{b110}.}
\label{aadfig4}
\end{figure}

If galaxies in groups and clusters merge with the central
galaxy over a finite time to produce a progressively greater massive central galaxy,
then one way to quantify the dynamical age of a galaxy system is to
measure the luminosity gap between the two most luminous galaxies
remaining in the group, provided that there is no infall of bright
galaxies from other nearby systems. A relaxed X-ray morphology in
observed fossils is an indication of the absence of such a process.
Within such a merging scenario for the formation of the brightest
group galaxy, the luminosity gap distribution depends on the halo
merger rates. This has  previously only been examined in analytical
studies \citep{b30}, or in large-scale cosmological simulations without
 a hot baryonic component. Thus, while 
an important observational criterion for defining fossils relies on the X-ray
properties of the group, this has not previously been 
implemented within the models.

\subsubsection{The R-band magnitude gap distribution}

Here we compare the luminosity gap distribution between the luminous
galaxies found at the centre of dark matter haloes extracted from the
Millennium simulation with expectations from the analytical model of
\citet{b115}, and with observational properties from SDSS clusters.

\citet{b115} compared the distribution of the predicted luminosity gaps 
from their analytical model within $R_{200}$, as a function of halo mass, 
with the observed 
luminosity gaps in the SDSS (DR4) clusters \citep{b110}, ranging in mass and
redshift from $M=0.5\!-\!10 \times 10^{14}\, h^{-1}\, $M$_{\odot}$ and
$z$=0.02 to 0.17, respectively, within a projected physical radius of 
500 $h^{-1}$kpc. 
Halo merger rates in their model have
been analytically estimated  according to the excursion-set theory of
\citet{b30}, which is also known as the extended Press-Schechter
formalism. Assuming a halo density profile of the form of \citet{b123}, a
subhalo of mass $m$ merges into a primary halo of mass $M$ ($m \geq
\frac{1}{2}M$) and makes a composite halo. As the centre of the 
subhalo crosses
the virial radius of the new composite halo, a merger happens. Then the
subhalo spirals toward the centre of the composite halo in a near circular
orbit, experiencing dynamical friction.

For comparison with the above study, we evaluate the $R$-band luminosity
difference between the first and second most luminous
($\Delta m_{12}$) and the first and the third most luminous
($\Delta m_{13}$) galaxies in our Millennium data
within $R_{200}$ and within 500 $h^{-1}$kpc. In Fig.~\ref{aadfig4}, we
plot the $R$-band luminosity gap
distribution of the Millennium simulation for the same mass range as the 
models, together with the luminosity
gap distribution of 730 SDSS C4 clusters  \citep{b110}.
 Figs.~\ref{aadfig4}a and ~\ref{aadfig4}b compare the predicted gap statistics 
from \citet{b115} for two values of the Coulomb
logarithm, $\ln \Lambda = 1$ and $\ln \Lambda = 2$,  within 
$R_{200}$. Since the parameter $\ln \Lambda$ is proportional to the force of dynamical friction 
between the  centres of subhalo and primary halo during the process of merging,  
a higher value of $\ln
\Lambda$ corresponds to a faster effective halo merger rate. In numerical
simulations, $\ln \Lambda$ is approximated by $b_{max}/b_{min}$, where
$b_{max}$ and $b_{min}$ are the maximum and minimum impact parameters
respectively, and $\ln \Lambda$ is expected to be $\sim 1-4$
\citep{b190,b57,b50}.  
However, in the semi-analytic galaxy
catalogues \citep{b40}, based on the Millennium simulation, used in
this work, the above relation is approximated by $ \ln \Lambda=\ln
(1+M_{\rm 200}/m_{\rm sat})$, where $m_{sat}$ is the halo mass of 
the satellite galaxy.

Within the mass range of the SDSS data there are 8842 haloes in the
Millennium simulation catalogue.  Accordingly, in Fig.~\ref{aadfig4},
 our data have been normalised to be comparable
with the SDSS data and the theoretical model of \citet{b115}. However, 
the simulation data is, unlike the
observations, complete and uncontaminated by spurious groups or
foreground and background galaxies. All these effects are likely to be
heavily dependent on the number of galaxies residing in the halo. As
such, the comparison with the SDSS data shown in Fig.~\ref{aadfig4} should
be treated with caution.

Given this caveat, our analysis based on the Millennium simulation
catalogues agrees remarkably well with the models of \citet{b115} based on
the SDSS survey for the luminosity gap distribution of the two brightest
galaxies in each of the dark matter haloes, particularly for $\ln \Lambda
\! =\! 2$ (Fig.~\ref{aadfig4}a). However, for the $R$-band luminosity gap
between the brightest and third brightest galaxies in each system
(Fig.~\ref{aadfig4}b), the simulations significantly depart from the model.
When comparing with the SDSS data (Figs.~\ref{aadfig4}c and
~\ref{aadfig4}d), the simulations overpredict the frequency of
groups. The simulations and the
SDSS data have similar shaped distributions for the luminosity gap 
$\Delta m_{13}$, but
with a shift of $\sim$0.5 mag toward higher $\Delta m_{13}$ in the
simulated haloes.

We emphasize that the Millennium predictions for the luminosity gap
statistic are sensitive to the assumed mass range and search
radius of dark matter haloes within which brightest halo
members are identified.  SDSS cluster masses have been 
estimated from total $r$-band luminosities, so any inaccuracies in
this procedure would affect the comparison with the Millennium data.

Observationally there is an excess population of groups with a
small luminosity gap between the first and second ranked
galaxies, above what is predicted by the theoretical models or the
simulations.  This excess population is likely to result from
contamination of observed group samples by local structure alignments,
and renormalising to a sample without these groups scales down
the ``Millennium'' distribution in  Fig.~\ref{aadfig4}c, bringing
the simulation results and the observational measurements into better
agreement.  Results are similar in the $K$-band.


\subsubsection{The abundance of fossil groups}
\label{abundance}

The probability of finding fossil systems is expected to
increase with decreasing halo mass, as shown in
previous studies based on theoretical models or hydrodynamical
simulations \citep{b50,b115,b147,b187}. Unfortunately, it is difficult to
compare the results from different studies (both theoretical and observational),
since they have used a range of search radii (from $R_{180}$ to $R_{337}$ --
see Table. 1) within which the $\Delta m_{12} \geq~2\,$mag criterion is 
imposed. Clearly, the larger the search radius, the more demanding is the
requirement on the galaxy contents of the system, and the smaller the fraction
of groups which will qualify as fossils.

\begin{figure}
\epsfig{file=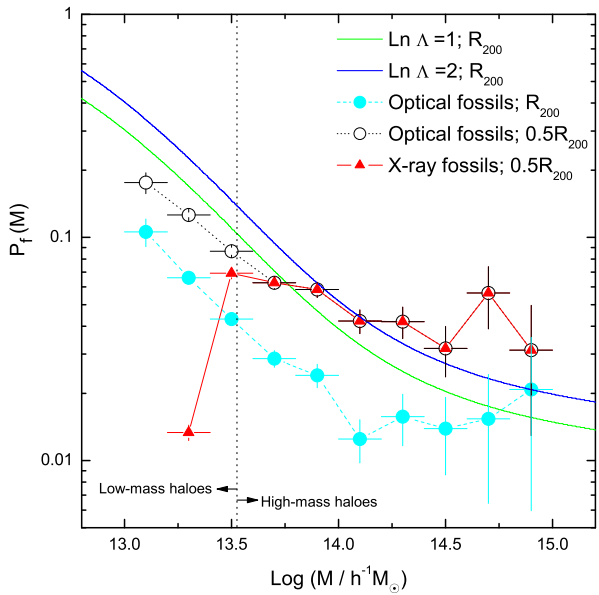,width=3.7in}
\caption[]{The probability, $P_f(M)$, that a dark matter halo of mass
$M$ contains an X-ray fossil group ({\it closed triangles}), 
optical fossil group within 0.5$R_{200}$ ({\it open circles}), 
or optical fossil group within $R_{200}$ ({\it closed circles}) from the Millennium
simulation. The fossil incidence
rate from the analytical study of \citet{b115} for two values of
Ln$\Lambda=1$ ({\it green line}) and Ln$\Lambda=2$ ({\it blue
line}) is also plotted. The vertical dotted-line corresponds to halo mass 
$\sim 3.34 \times 10^{13}\,h^{-1}\,$M$_{\odot}$ (see Sec.~\ref{evolution}).}
\label{aadfig5}
\end{figure}

In Fig.~\ref{aadfig5},
the rates of incidence, $P_f(M)$, of optical fossils and X-ray fossils
(using our preferred search radius of 0.5$R_{200}$, following  \citet{b65})
are plotted, as a function of the mass $M$ of the halo, together with
the predicted values from the models of
\citet{b115} for two values of $\Lambda$. The shape of our curve for
optical fossils is quite similar to the theoretical models (which included
no X-ray luminosity criterion), but the latter actually employed a search
radius of $R_{200}$. To see the effect of this, we also show our Millennium
results for this larger search radius. The fraction of fossil systems falls
by approximately a factor of 2, when this more demanding requirement is imposed,
and so lies significantly below that predicted by \citet{b115}.

On scales of $M \!\sim\!
10^{13}\!-\!10^{14}h^{-1}$M$_{\odot}$, $\sim$5\%--18\% of groups are
optical fossils. This probability falls to $\sim$3\%--5\% for more
massive ($M\geq 10^{14}h^{-1}$M$_{\odot}$) fossil systems.
For halo masses $>5\times10^{13}h^{-1}$M$_{\odot}$ all optical fossils
in the simulation are also X-ray fossils. However, at
the lowest halo masses the fraction of X-ray fossils drops steeply, 
since many low mass haloes do not satisfy the $L_X$ threshold criterion. 

In Table. 1, we summarize the incidence rates of fossil systems from
present study as well as those found in the literature. Comparison between
these different estimates is difficult, since both the search
radius and the halo mass range varies considerably from study to
study. However, a direct comparison with the only {\it observational} estimate
(from \citet{b65}) is possible, since we have used the same definitions
of fossil groups as these authors.
Based on a comparison with the integrated local X-ray luminosity function
of \citet{b56}, \citet{b65} estimated that X-ray fossil systems constitute
8-20\% of all systems of the same X-ray luminosity ($L_{X,bol} \geq
0.25\times 10^{42} h^{-2}$erg s$^{-1}$).  The right panel histogram of
Fig.~\ref{aadfig2} represents the fraction of optical fossil systems in
each bin of $L_{X,bol}$. Integrating this over all X-ray luminosities
above the threshold value for fossils, we find 
that $\sim 7.2\pm 0.2\%$ of haloes with
$L_{X,bol} \geq 0.25\times 10^{42} h^{-2}$erg s$^{-1}$ are X-ray fossils,
which is reasonably consistent with the lower limit of $\sim 8\%$,
derived by \citet{b65}.

In comparison, detailed hydrodynamical
simulations by \citet{b50} and \citet{b147} of 12 galaxy groups,
predict a larger fraction of 33\%$\pm$16\% for fossil systems of mass
$10^{14}h^{-1}$M$_{\odot}$ or larger. This may be because it is easy
to overestimate the local viscosity in hydrodynamic simulations
\citep{b166}, a process that would lead to central overmerging in the
models. 

\begin{table*}
\label{aabundance}
\centering
\begin{minipage}{140mm}
\caption{The incidence rates of fossil systems.}
\begin{tabular}{@{}llclll@{}}
\hline
  
Mass range  ($h^{-1}$M$_{\odot}$)          & $L_X$ ($10^{42}h^{-2}$ erg s$^{-1}$)        & Fossil type 
\footnote{O: Optical fossils , X: X-ray fossils.} & Search radius &  Fossil fraction (\%)& Reference\footnote{S07:~\citet{b142}; M06:~\citet{b115}; vdB07:~\citet{b187}; SL06:~\citet{b147}; DO05:~\citet{b50}; J03:~\citet{b65}.} \\ 
\hline
$\!\sim\! 10^{13}\!-\!10^{14}$  &       -     & O &      $R_{200}$      & $\sim 5-40$        & M06     \\
$\!\sim\! 10^{13}\!-\!10^{14}$  &       -     & O &      $R_{180}$      & $\sim 3.6 \pm 0.1$ & vdB07\footnote{Based on the conditional luminosity function (CLF) formalism of \citet{b187}.}   \\
$\geq 10^{14}$                  &       -     & O &      $R_{200}$      & $\sim 1-3$         & M06     \\
$\!\sim\! 10^{14}\!-\!10^{15}$  &       -     & O &      $R_{180}$      & $\sim 6.5 \pm 0.1$ & vdB07   \\
$\!\sim\! 10^{14}$              &       -     & O &      $R_{337}$      & $\sim 33 \pm 16$   & SL06, DO05\footnote{From hydrodynamical simulations of 12 galaxy groups.}     \\
$\!\sim\! 10^{13}\!-\!10^{15}$  & -           & O &  1$~h^{-1}$Mpc      & $\sim 8-10$        & S07\footnote{Based on the Millennium simulation. The  first brightest galaxies of fossils in their sample are  always brighter than $M_R=-20.5$.}     \\ 
 -                              & $\geq 0.25$ & X &   0.5$R_{200}$      & $\sim 8-20$        & J03     \\ 
 -                              & $\geq 0.25$ & X &   0.5$R_{200}$      & $\sim 7.2 \pm 0.2$ & Present study\footnote{Histogram on the right panel of Fig.~\ref{aadfig2}, gives the fraction of X-ray and optical fossils in each bin of $L_X$.} \\
$\!\sim\! 10^{13}\!-\!10^{15}$  & -           & O &   0.5$R_{200}$      & $\sim 13.3 \pm 0.2$& Present study \\ \hline

\end{tabular}
\end{minipage}
\end{table*}
 
\subsection{The Space Density of X-ray Fossil Groups}

So far, the integrated space density of X-ray fossil groups has been studied
for small samples, each of three to five X-ray fossil systems, at different
limiting luminosities \citep{b170,b140,b65}. Here, we estimate the
space density by systematically counting the fossil groups in the
whole 500 $h^{-1}$ Mpc survey volume of the Millennium Simulation at
$z\!=\!0$.  For comparison with previous studies, we select and count
X-ray fossil groups for three limiting X-ray luminosities ranging from
0.25--5$\times 10^{42}$ $h^{-2}$erg s$^{-1}$.  The space densities calculated
 at different limiting luminosities as well as those from previous
studies are given in Table~2. The value from Romer et al is a very rough
estimate, since no redshifts for galaxies surrounding the central
object were available in this study.

Our values show that for X-ray luminosities exceeding 
2.5-5$\times 10^{42}$ $h^{-2}$erg s$^{-1}$,
the space density of fossils in the Millennium simulation agrees within 
the errors with those estimated by \citet{b170} and \citet{b65}. 
At the lowest X-ray luminosities, the density from the Millennium simulations 
appear 
to be lower than observed, though the observational values given in
Table~2 have large uncertainties due to the small number of
X-ray fossil groups and the effects of cosmic variance. Recent studies of
\citet{b85} and \citet{b59} show that one of the fossils in the
sample of \citet{b65} does not satisfy the fossil criterion of $\Delta
m_{12}\geq 2$, which reduces the observational space
density.   Certainly the number of X-ray fossils found is heavily
dependent on the X-ray luminosity threshold chosen and may be
influenced by the scatter in X-ray group properties near this lower
limit.

\begin{table}
\label{density}
\centering
\begin{minipage}{140mm}
\caption{Space densities of fossil galaxy groups.}
\begin{tabular}{@{}lcllll@{}}
\hline
  
${L_X}$\footnote{In units of $ 10^{42}h^{-2}$ erg s$^{-1}$}& ${N_f}$\footnote{Number of fossils}& Density\footnote{In units of $ 10^{-7}h^3$ Mpc$^{-3}$ }            & Reference\footnote{V99:\citet{b170}, R00:\citet{b140}, J03:\citet{b65}} & Present study$^c
$ \\ \hline
 $>0.25$ & 5        & $320^{+216}_{-144}   $ & J03           &  $104  \pm 3  $   \\
 $>2.5 $ & 3        & $16^{+15.2}_{-8.8}   $ & J03           &  $22.4 \pm 1.3$   \\
 $>2.5 $ & 4        & $36.8^{+47.2}_{-18.4}$ & V99           &  $22.4 \pm 1.3$   \\
 $>2.5 $ & 3        & $\sim 160            $ & R00           &  $22.4 \pm 1.3$   \\
 $>5.0 $ & 4        & $19.2^{+24.8}_{-9.6} $ & V99           &  $12.8 \pm 1.0$   \\ 
\hline
\end{tabular}
\end{minipage}
\end{table}


\begin{figure*}
\centering 
\epsfig{file=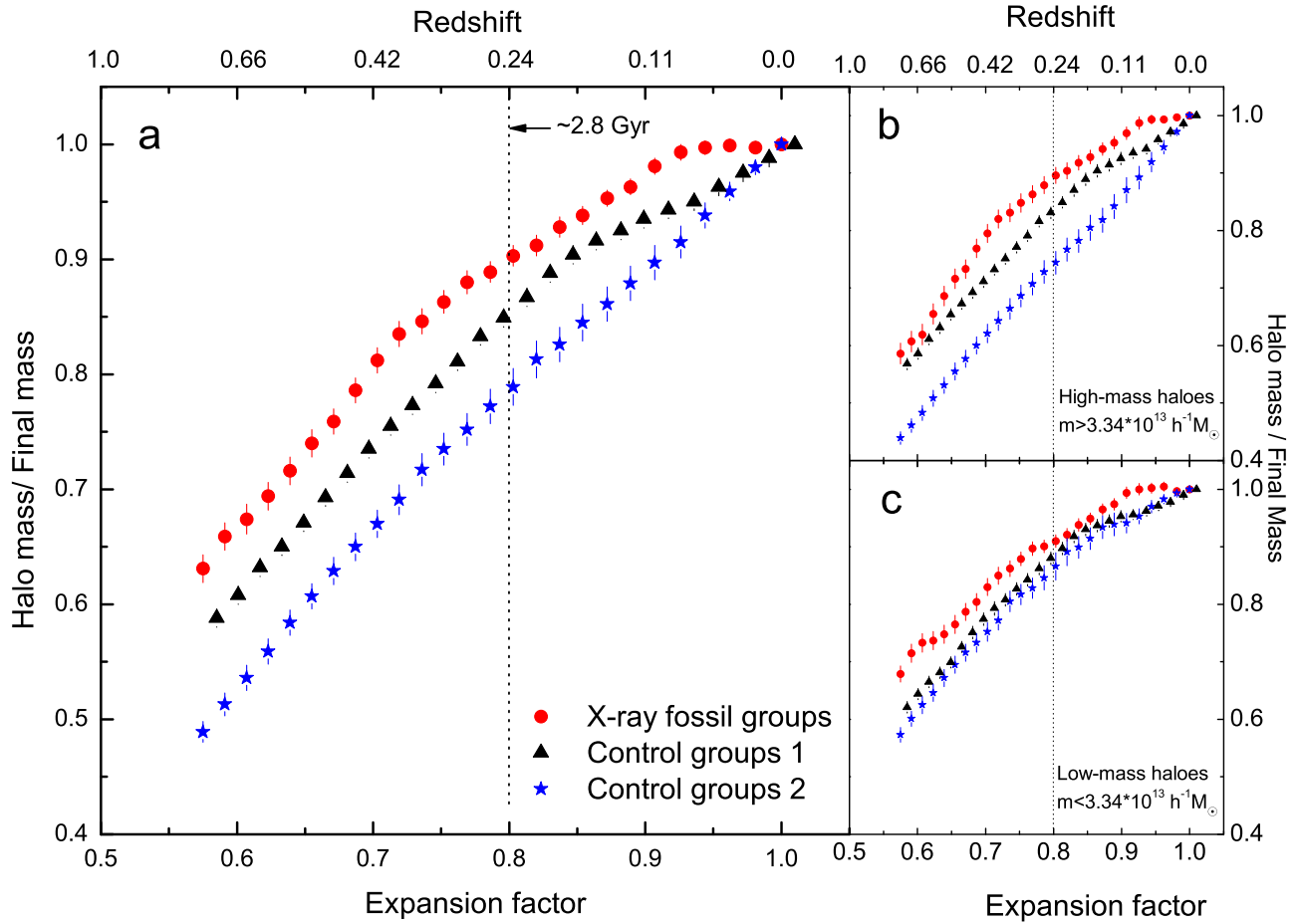,width=7.4in}
\caption[] {Tracing back the mass build-up of the dark matter haloes
as a function of expansion factor and redshift for both the X-ray fossils and
the control groups 1 and 2.  (a) For all haloes. (b) High mass haloes ($m_{\rm halo}\gtrsim 3.34\times
10^{13}\,h^{-1}$ M$_{\odot}$). Both plots indicate the earlier formation of X-ray fossil groups in comparison to
control groups. (c) Low mass haloes ($m_{\rm halo}\lesssim
3.34\times 10^{13}\,h^{-1}$ M$_{\odot}$). Here the difference
in evolution between the X-ray fossil and control groups is not as pronounced as in those seen in high mass X-ray fossils. 
All the masses are normalised to the mass at $z$=0.}
\label{aadfig7}
\end{figure*}

\subsection{Evolution of Fossil Groups}
\label{evolution}
 
Strong interactions and mergers between galaxies occur more
efficiently in the low velocity dispersion environment of galaxy
groups \citep{b105}.  Therefore in old, relatively isolated groups, most massive
galaxies have sufficient time to merge via dynamical friction. If
X-ray fossil groups are indeed systems that formed at an earlier epoch, we
should be able to verify this from the merger histories of present-day
fossils in the Millennium simulation: an exercise that is not directly
possible to perform with observational surveys. In Fig.~\ref{aadfig7}
we trace the mass evolution of present-day X-ray fossil systems backwards
from $z\!=\!0$, to $z\!=\!0.82$ when the scale factor, $a$, of the
Universe was 0.55 times its current size.
 
At any given redshift in Fig.~\ref{aadfig7} the average ratio of the
mass of a halo to its final mass (at $z\!=\!0$) is calculated for all
eligible haloes.  The error is represented by the standard error on
the mean, i.e. $\sigma/\sqrt{N}$, where $\sigma$ is the standard
deviation of the original distribution and $N$ is the sample size. The
same was done for both sets of control groups. The original sample of
fossils was divided into two subsamples, of low mass and high mass
groups (see Fig.~\ref{aadfig7}b and Fig.~\ref{aadfig7}c), such that both subsamples have equal numbers of groups.
The boundary between the two subsamples
corresponds to the median present-day mass  
$\sim 3.34 \times 10^{13}\,h^{-1}\,$M$_{\odot}$.

Fig.~\ref{aadfig7}a  shows that at a scale factor of 0.8 ($z\sim 0.24$), the
 fossil groups have already attained $\sim$90\% of their
final mass while, at the same redshift, the fraction of assembled
mass of the extreme non-fossil groups is about $\sim$77\% of their
final mass.  The intermediate control group gives intermediate
values. The fossil groups have almost all their mass in place by a
redshift of $z\! \sim\! 0.1$, and show no evidence of  recent major
mergers, while the non-fossils seem to be assembling mass even at the
present day. These results suggest an early formation and consequent
higher mass concentration in fossil groups, in comparison to normal
groups, particularly for the more massive fossils.

As Figs.~\ref{aadfig7}b and ~\ref{aadfig7}c show, the
difference in mass assembly is larger in more massive haloes than haloes with lower mass. 
The decreased distinction in the assembly history for our lower mass fossil systems
probably results from the fact there is a large fraction of 
``statistical fossils'' in this category: groups which achieve  $\Delta m_{12} \geq 2\,$
due to random chance, because of the small number of members. As can be seen in
Fig.~\ref{aadfig3}, $\sim$50\% of  optical fossil groups with
masses less than $\sim 3.3 \times 10^{13}\,h^{-1}\,$M$_{\odot}$  
are expected to fall into this ``statistical fossil'' category.

Various observational properties \citep{b130,b65,b75,b167,b80,b85} have suggested an
early formation epoch for fossils. \citet{b50} and \citet{b147} used a set of twelve
high-resolution numerical simulations in the $\Lambda$CDM cosmology to
study the formation of fossil groups, and found a correlation for the
magnitude gap between the brightest and second-brightest
galaxies and the halo formation epoch, with fossils accreting half of
their final dark matter mass at $z\geq 1$. Such an early assembly of
fossil haloes leaves enough time for L$_{\ast}$ galaxies to merge into the
central one by dynamical friction, resulting in the observed magnitude gap at
$z=0$.
 
\begin{figure}
\begin{center}
\epsfig{file=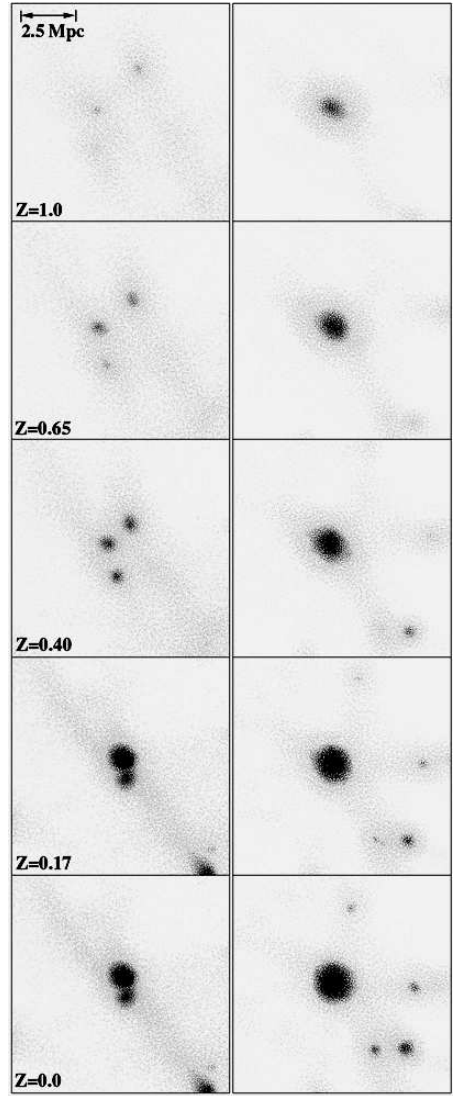,height=5.75in}
\caption[]{Evolution of a typical of massive X-ray fossil group ({\it
right}) in comparison to a typical massive normal group ({\it
left}) from redshift $z$=1.0 to 0. The dimension of each panel is $10
\times 10$ Mpc. The points represent individual gas particles from the
Millennium Gas simulation \citep{b128}.}
\label{aadfig9}
\end{center}
\end{figure}

 Fig.~\ref{aadfig9} shows the history of mass assembly of a typical example of a massive
fossil group (right panel) and a control group (left panel) from the
Millennium Gas Simulation from redshift $z=1.0$ to $z=0$. The
dimension of each image is $10 \times 10$ Mpc, centred around the
central halo.  It can be seen that at $z=0.3$, the X-ray fossil group has
already largely been assembled, while the control group has
considerable substructure even at a later epoch. 


\section{Discussion}

\begin{figure}
\begin{center}
\epsfig{file=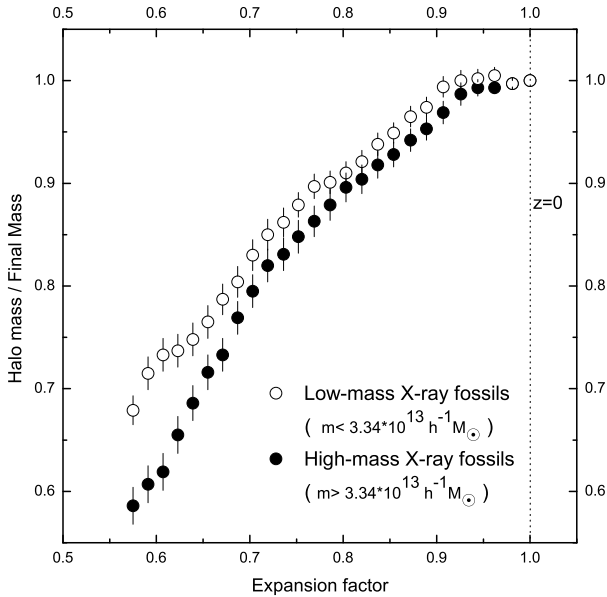,width=3.7in}
\caption[]{Comparison between the mass growth of high and low-mass X-ray fossils plotted in Figs.~\ref{aadfig7}b and ~\ref{aadfig7}c.}
\label{aadfig10}
\end{center}
\end{figure}

We studied the history of the mass assembly of fossil groups, selected
using the usual observational criteria at $z\!=\!0$, from a redshift
$z\!=\!0.8$ to the present day, within the Millennium simulation. 
A sample of X-ray fossil groups was defined from the Millennium
simulations and associated gas and galaxy catalogues, according to the
usual criteria: (a) the difference between the $R$-band magnitudes of the
first and second ranked galaxies, within half the projected radius
enclosing 200 times the mean density of material ($R_{200}$), is
$\Delta m_{12}\!\geq\! 2$ magnitudes, and (b) The bolometric X-ray luminosity of the group is 
$L_{\rm X,bol} \geq 0.25 \times 10^{42} h^{-2}$erg
s$^{-1}$. While optical fossil groups fulfill just the first condition, X-ray fossils satisfy both criteria.
Our main results are as follows:

\begin{itemize}

\item The space density of X-ray fossil groups is in close agreement with the
observed space density of fossils with $L_X>2.5\times 10^{42}$ $h^{-2}$erg s$^{-1}$. 
Although for low luminosity fossils we find roughly 1/3 of the
observed fossil space density, there are several potential
factors that could lead to this difference. As well as
significant uncertainties in the observational studies, the
X-ray properties of haloes in the real Universe show far greater scatter than
those seen in the preheating simulation used here \citep{Hartley}.
Given the X-ray luminosity threshold in the definition of an X-ray
fossil, scatter in $L_X$ will
alter the X-ray fossil number density, since the number density of
haloes is a steep function of mass.

\item By selecting optical fossils from groups randomly generated from a
Schechter luminosity function, we
demonstrate that for small numbers of galaxies per group, a significant fraction
of optical fossil groups are expected to be purely statistical, requiring no special
physical mechanism to generate the 2 magnitude luminosity gap.
 For groups with more than 40 members,
this effect largely disappears, with very few  fossil groups
expected at random. 

\item The probability of finding optical fossils with mass $M$, i.e, 
$P_f(M; \hbox{optical})$ is a decreasing function of: (a) group dark matter halo mass 
and, (b) the fraction of
the virial radius within which the first and second brightest galaxies
are being found.  Conversely, as dark matter halo mass become small,
the probability $P_f(M; \hbox{X-ray})$ for X-ray fossils decreases.

\item Both high-mass and low-mass X-ray fossil groups are found to have assembled $\sim 90\%$
of their final masses by a redshift of $z\!=\!0.24$. The corresponding
mass fraction is about $\approx 70-80\%$ for two different sets of
high-mass control samples, and $\approx 85\%$ for low-mass control samples, 
where groups fulfill the same X-ray luminosity
criterion ($\gtrsim 3.34 \times 10^{14}\,h^{-1}$M$_{\odot}$) but have the optical luminosity gaps corresponding 
$0.1 \leq\! \Delta m_{12}\!\leq\! 0.3$ and $0.8 \leq\! \Delta
m_{12}\!\leq\! 1.0$~magnitudes. 



\end{itemize}
   
This study shows that fossils indeed are formed early, with more than 
$\sim$80\%  of
their mass accumulated as early as 4~Gyr ago. They are also relatively 
isolated compared to non-fossils. The strongest X-ray fossil
candidates are those with the highest X-ray luminosity as these
systems are not expected to have a large luminosity gap between their
first and second ranked galaxies entirely by chance. As always,
systems with more than a handful of galaxies are to be preferred.

In principle, comparison of the observed space density of fossils as a function
of X-ray luminosity with that of fossils from simulations can provide valuable
constraints on the treatment of physical processes included in the simulations.
The tentative evidence for a discrepancy, whereby the observed space density
of low X-ray luminosity fossil groups
may exceed that predicted from the simulations, will be worth revisiting in
the future, when better observational estimates are available.

It is interesting that while the amount of recent mass
assembly in control groups increases with halo mass, as is expected in the 
hierarchical growth paradigm,
there is almost no difference between the mass assembly of high-mass and 
low-mass X-ray fossils after redshift $z=0.6$  (see Fig.~\ref{aadfig10}). 
It seems that both low-mass and high-mass fossil systems are undisturbed
at low redshift.

Since we expect faster orbital decay and more efficient galaxy merging in
lower mass systems, due to the lower velocity dispersion of individual
galaxies within the group, we would expect to find a higher incidence rate
of fossils amongst poor groups. Fig.\ref{aadfig3} shows that this is indeed
the case, but the effect is not very strong, once the influence of
statistical fossils is removed, and in X-ray fossils any rising trend
at low richness is overwhelmed by the fact that many of the optical
fossils fail to exceed the X-ray luminosity threshold.
A word of caution about the treatment of orbital decay is in order here. As
discussed in Sec.~\ref{semianalytic}, the orbital evolution of subhaloes 
in the Millennium
simulation is well treated until these subhaloes are reduced by stripping to
20 dark matter particles, but thereafter is calculated semi-analytically,
using an approximate formula. In practice, a significant fraction of the
second ranked galaxies in the fossil systems we have extracted from the
simulation have been stripped below this 20 particle limit.
For example, in fossils with masses of only $10^{13}\,h^{-1}\,$M$_{\odot}$, approximately
35\% of second ranked galaxies have been stripped below the limit,
though this fraction drops to $\sim$10\% for systems 
with mass $> 10^{14}\,h^{-1}\,$M$_{\odot}$. For such galaxies the timescale 
for their subsequent decay and merger with the central galaxy is not 
very reliable. However, this is not a major issue for massive haloes, and
it is interesting and surprising that the incidence of fossils in 
rich systems is fairly flat at 3-4\%. Observational studies should,
in due course, show whether this is reflected in the real Universe.
One example of a fairly rich fossil cluster has already been reported by
\citet{b80}.

The magnitude gap distribution of haloes at different X-ray 
luminosities and the mass evolution of fossil groups discussed above both
support the idea that X-ray fossil groups are not a distinct class of objects
but rather that they are extreme examples of groups which collapse early
and experience little recent growth, so that their galaxies have time
to undergo orbital decay and merging. The X-ray and optical scaling properties
of such extreme groups can be expected to differ from those of groups
with more typical evolutionary histories, and such differences have already
been observed \citep{b85}. A comparison of such observed differences
with the properties seen in the Millennium simulation groups is underway,
and should provide a valuable check on the adequacy with which feedback 
processes and other baryon physics is handled in the simulations.

\section*{Acknowledgments}
The Millennium Simulation used in this paper was carried out by the
Virgo Supercomputing Consortium at the Computing Centre of the
Max-Planck Society in Garching. The semi-analytic galaxy catalogue is
publicly available at http://www.mpa-garching.mpg.de/galform/agnpaper. 

The Millennium Gas Simulations were carried out at the Nottingham
HPC facility, as was much of the analysis required by this work.
We would like to thank Arif Babul and Serena Bertone for the discussions
 on this topic.

\label{lastpage}

\end{document}